# Security–efficiency optimization in multi-threshold intensity-modulated discrete-variable secret-key distribution

AMIR YAZDANPOUR,* SHIVA KUMAR, NAHID SHARIFI, AND YEGANEH NASR

McMaster University, Electrical Engineering Department, 1280 Main Street West, Hamilton, ON L8S 4L8, Canada
*Yazdaa8@McMaster.ca

**This paper presents a new discrete-variable secret-key distribution (DV-SKD) protocol based on multi-threshold (MT) intensity-modulation and direct-detection (IMDD). The protocol employs a multi-symbol decision structure with two erasure regions and a joint optimization of modulation depths and decision thresholds, enabling enhanced control over the trade-off between symbol distinguishability and secrecy. This multi-threshold architecture increases the information extracted per accepted detection event while preserving the simplicity of IMDD hardware. We develop a full optimization framework for the modulation parameters under realistic noise models and evaluate performance using both forward and reverse reconciliation. Numerical simulations show that MT-IMDD achieves robust key-generation rates (KGR) under practical operating conditions and maintains strong resilience against both passive and active eavesdropping attacks.**

## 1. INTRODUCTION

Quantum cryptography, and in particular, quantum key distribution (QKD), is widely regarded as the heart of future cybersecurity. Owing to its foundation in quantum mechanics, QKD offers information-theoretic security with rigorous theoretical proofs, making it a highly trustworthy approach. Nevertheless, significant opportunities remain to enhance its practical performance and implementation. Based on the characteristics of the transmitter (Alice) and receiver (Bob), QKD protocols are generally categorized into three families: continuous-variable (CV) QKD, discrete-variable (DV) QKD, and hybrid QKD. The very first protocol developed in 1984 by Bennett and Brassard [1] is a DV-QKD protocol. DV-QKD uses discrete degrees of freedom such as polarization [1–7]. CV-QKD encodes information in field quadratures and is typically measured with homodyne or heterodyne detection [8–14]. Hybrid protocols mixing continuous encodings with photon counting have also been proposed and experimentally demonstrated [15–20].

While CV and hybrid schemes offer strong security guarantees, they often require phase-stable local oscillators and careful calibration, whereas DV-QKD, though capable of long-distance key distribution, relies on single-photon sources or weak coherent pulses and avalanche photodiodes (APDs), which are widely used in practical systems, or superconducting nanowire single-photon detectors (SNSPDs), which offer superior performance at the expense of cryogenic operation. Typically, CV-QKD relies on balanced photodiodes and coherent detection, which introduces additional complexity in phase referencing and local oscillator (LO) management. By contrast, intensity-modulation with direct-detection (IMDD) can leverage mature components and existing fiber-optic infrastructure, enabling simpler and more cost-effective implementations [21]. Prior IMDD–based Secret Key Distribution (SKD) schemes [21,22] have demonstrated that meaningful secrecy rates are achievable with commercially available hardware under realistic receiver-noise conditions.

In this work, we propose a new IMDD–based SKD protocol, which we refer to as the multi-threshold IMDD (MT-IMDD) scheme. It employs a multi-symbol intensity alphabet together with multiple decision thresholds and erasure regions, enabling flexible control over the information–secrecy trade-off. The resulting decision structure allows the receiver to extract more information per accepted symbol while suppressing Eve's distinguishability through tunable decision margins. We develop a joint optimization framework for the modulation depths and threshold locations under realistic noise conditions, showing that the architecture provides favorable symbol distinguishability and strong resilience against eavesdropping while preserving hardware simplicity. Although the model does not provide unconditional security, it represents a compelling and practical approach to high-rate DV-SKD over existing fiber links. To assess performance, we evaluate the proposed protocol under two representative eavesdropping strategies, beam-splitting attack (BSA) and intercept–resend attack (IRA), using both forward and reverse reconciliation.

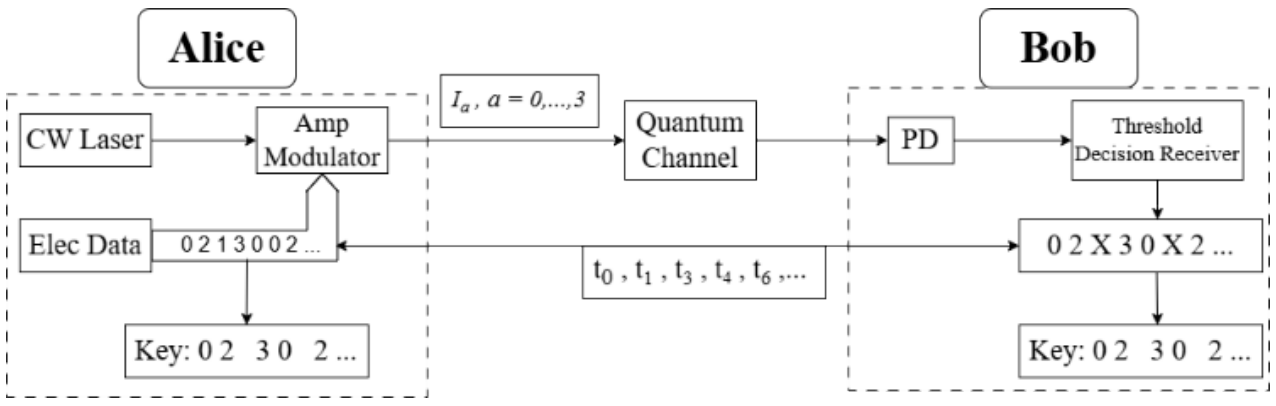

**Fig. 1.** Schematic of the MT-IMDD SKD protocol. Amp Modulator, amplitude modulator; CW Laser, continuous-wave laser; Elec Data, electronic data; PD, photodetector; $t_0$, $t_1$, …, time slots used for symbol transmission. Bob uses a threshold-based decision receiver to recover the transmitted symbols and reconstruct the key.

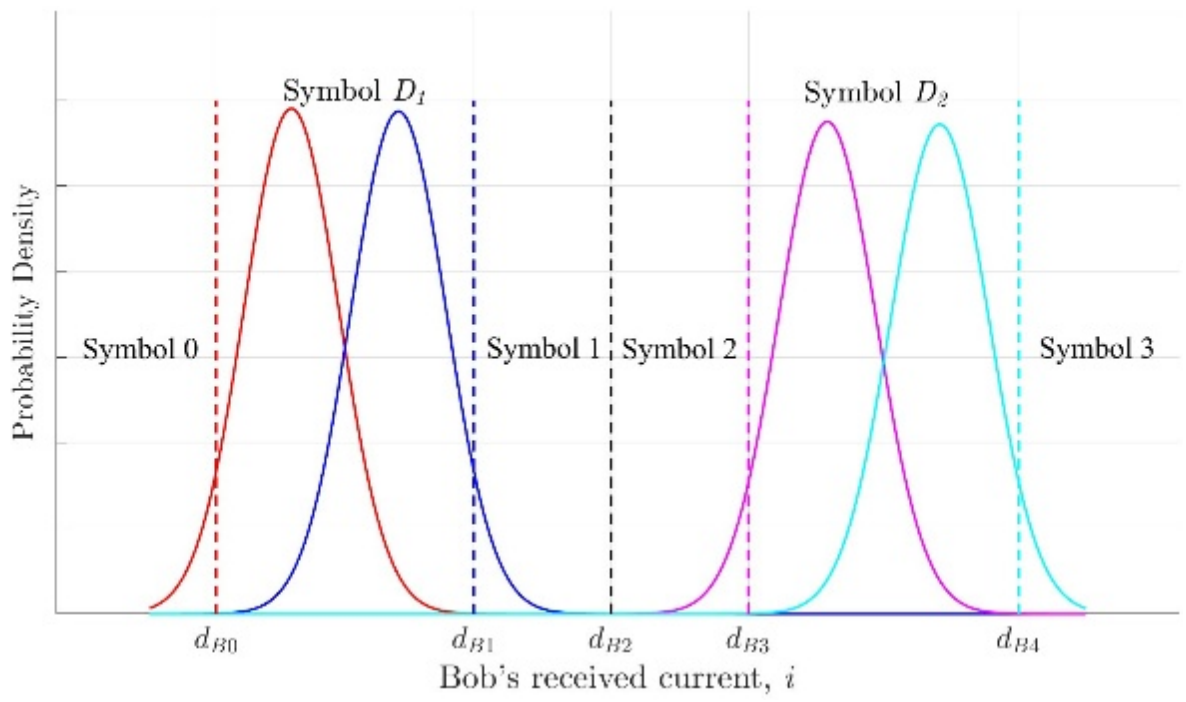

**Fig. 2.** Bob's received-current probability densities for the four intensity-modulated symbols in the proposed multi-threshold IMDD (MT-IMDD) protocol. Solid curves show the conditional Gaussian distributions corresponding to each transmitted intensity level. The dashed vertical lines denote Bob's decision thresholds $d_{B0}, d_{B1}, d_{B2}, d_{B3}, d_{B4}$, which partition the measurement axis into symbol regions $\{0, 1, 2, 3\}$ and two erasure regions $D_1$ and $D_2$.

The paper is organized as follows. Section 2 reviews IMDD and its integration into the proposed framework (MT-IMDD). Section 3 develops the mutual-information and key-rate formulations with sifting. Sections 4 and 5 evaluate BSA and IRA, respectively, under these assumptions.

## 2. MT-IMDD PROTOCOL

Figure 1 shows a schematic of IMDD SKD system based on MT. Alice sends one of the intensity levels $\{I_0, I_1, I_2, I_3\}$ over a symbol interval to transmit the symbols $\{0, 1, 2, 3\}$, respectively. Intensity level when the symbol $a$ transmitted is:

$$I_a = I_0(1 + \delta_a), \ a = 1, 2, 3, \quad (1)$$

where $I_0$ is the mean intensity of symbol 0, and $\delta_a \ll 1$. As shown in Fig. 2, Bob sets 5 thresholds $d_{B,j}$, $j = 0, 1, ..., 4$. When the measured current is smaller than $d_{B,0}$, he creates symbol 0 and when the measured current is between $d_{B,1}$ and $d_{B,2}$, he creates symbol 1. However, if the measured current, $i$, is greater than $d_{B,0}$ but still less than $d_{B,1}$, he creates no symbol (symbol $D_1$). If the received current is in the range $[d_{B,2}, d_{B,3}]$, he creates symbol 2 (see Fig. 2). Finally, if the current is greater than $d_{B,4}$, symbol 3 will be created. Furthermore, if the current is in the range $[d_{B,3}, d_{B,4}]$, no symbol will be created (symbol $D_2$). A large $\delta_a$ results in well-separated intensity levels, lowering the symbol error rate (SER) since Bob can distinguish between symbols. However, intensity levels become more distinguishable for Eve, too, reducing the security of the system. Conversely, a small creates an overlapping intensity distribution, increasing the SER but making Eve's attack more difficult, thereby improving security. The optimal $\delta_a s$ are determined numerically based on the target transmission distance, noise levels, and eavesdropping risk.

By defining an "erasure" or "neutral" zones (symbols $D_1$ and $D_2$ in Fig. 2), Bob simply discards any measurement that falls into these ambiguous regions. This ensures that the symbols he does keep have a very high confidence level, significantly lowering the QBER of the raw key. If Eve tries to intercept the symbols, her own measurement will often fall into the erasure regions. If she is forced to "guess" and resend a symbol to Bob, she will introduce errors that are statistically visible. Therefore, erasure regions improve the reliability of the accepted symbols without providing an equivalent advantage to Eve. In addition, this selective discarding benefits the legitimate receiver (Bob) through the sifting process, as it reduces the contribution of low-confidence decisions to the key. Eve, however, cannot exploit this mechanism in the same way. In the beam-splitting attack, Eve passively observes the signal and does not control the sifting process. In the proposed multi-level setting, overlaps occur between multiple adjacent symbol pairs, which naturally give rise to multiple erasure regions ($D_1$ and $D_2$).

The security of this protocol relies on the fact that coherent states with a small amplitude difference are non-orthogonal to each other. Hence, an eavesdropper, Eve, cannot fully distinguish the transmitted state. The quantum states Alice transmits can be described as coherent states $|\alpha\rangle$, which are the eigenstates of the annihilation operator. The general form of a coherent state in the photon number basis is:

$$|\alpha\rangle = e^{-\frac{|\alpha|^2}{2}} \sum_{n=0}^{\infty} \frac{\alpha^n}{\sqrt{n!}} |n\rangle, \quad (2)$$

where $\alpha$ is a complex amplitude related to the intensity of the optical field. In an IMDD system, intensity levels correspond to different values of $|\alpha|^2$, since the mean photon number of a coherent state is given by $\langle n\rangle = |\alpha|^2$. The absolute square of the inner product between two coherent states is [23]:

$$\left\langle \alpha_j | \alpha_{j+1} \right\rangle = e^{-|\alpha_j - \alpha_{j+1}|^2}. \quad (3)$$

Since the intensity $I_j \propto |\alpha_j|^2$, it follows that a larger intensity difference $|I_j - I_{j+1}|$ results in a lower overlap, making the states more distinguishable. If Eve attempts a beam-splitting attack, she intercepts a fraction of the transmitted signal. However, due to the uncorrelated quantum noise in her measurement, her estimated key bits will not perfectly match Bob's, introducing errors in her observation. In an intercept-resend attack, she cannot perfectly distinguish the intensity levels due to their nonorthogonality.

## 3. MUTUAL INFORMATION AND KEY GENERATION RATE

In this section, two important parameters, mutual information (MI) and key generation rate (KGR), will be discussed. These parameters are important because they show how the schemes can perform under different situations.

### A. Mutual Information

In any key distribution system, noise arising from thermal, quantum, and classical sources degrades the received signal quality and limits the amount of shared information between the legitimate parties, Alice and Bob. In the present discrete-variable (DV) protocol, Alice transmits one of four discrete intensity levels, while Bob (or Eve) performs threshold-based detection that maps the continuous photocurrent into discrete symbol decisions. Consequently, the mutual information between two parties$X, Y = \{Alice, Bob, Eve\}$ is evaluated using the discrete form of Shannon's definition [24]:

$$I(X;Y) = \sum_x \sum_y P_{X,Y}(x,y) \log_2 \left[ \frac{P_{X,Y}(x,y)}{P_X(x) P_Y(y)} \right], \quad (4)$$

where $P_{X,Y}(x,y)$denotes the joint probability that symbols $x$ and $y$ occur, and$P_X(x)$, $P_Y(y)$ are the marginal probability densities of $X$ and $Y$, respectively. In our model, these discrete probabilities are obtained by integrating the underlying Gaussian conditional probability density functions over the corresponding decision regions defined by the detection thresholds at Bob's or Eve's receiver. This formulation connects the continuous optical noise model to the discrete-symbol statistics used for key-rate evaluation.

### B. Mutual Information between Alice and Bob

We first consider the joint probabilities between Alice and Bob. The conditional probability density function (pdf) of the received current $i$ when Alice sends symbol $a$ of mean intensity $I_a$ is:

$$p_{B|A}(b = i|a) = \frac{1}{\sqrt{2\pi}\sigma_a} \exp\left[ -\frac{(i - i_a)^2}{2\sigma_a^2} \right], \quad (5)$$

where

$$i_a = \kappa T \eta I_a. \quad (6)$$

$$\begin{aligned} \sigma_a^2 &= \kappa^2(\sigma_{C_a}^2 + \sigma_{Q_a}^2) + \sigma_T^2, \ a = 0, 1, ..., 3 \\ \sigma_{C_a}^2 &= k_1 B T^2 \eta^2 I_a^2, \\ \sigma_{Q_a}^2 &= k_2 B T \eta I_a, \\ \sigma_T^2 &= k_3 B, \end{aligned} \quad (7)$$

where $B$ is the bandwidth of the receiver, $T$ is the fiber transmissivity, $\eta$ is the quantum efficiency of the detector, $k_1$, $k_2$, and $k_3$ are the proportionality constants for the classical, quantum, and thermal noises, respectively. The proportionality factor $\kappa = q/(hf)$ relates the optical intensity to the generated photocurrent, where $q$ is the elementary charge ($1.602 \times 10^{-19}$ C), $h$ is Planck's constant ($6.626 \times 10^{-34}$ J · s), and $f$ is the optical carrier frequency ($f = c/\lambda$). Bob performs symbol decisions by comparing the current $i$ to the decision thresholds $\{d_{B0}, d_{B1}, d_{B2}, d_{B3}, d_{B4}\}$, which partition the measurement axis into six disjoint decision regions corresponding to the outcomes $b \in \{0, D_1, 1, 2, D_2, 3\}$(see Fig. 2). The conditional probability that Bob detects a symbol $b = b_j$ when Alice transmits symbol $a$ is then obtained by integrating the conditional pdf of $i$ over the appropriate decision region. The conditional probability is:

$$P(b_j|a) = \Pr\left\{ d_{B,j-1} \le i \le d_{B,j} | a \right\} = \Phi\left( \frac{d_{B,j} - i_a}{\sigma_a} \right) - \Phi\left( \frac{d_{B,j-1} - i_a}{\sigma_a} \right), \ j = 0, 1, ..., 5 \quad (8)$$

where $d_{B,-1} = -\infty$, $d_{B,5} = +\infty$, and $\Phi$is cumulative distribution function defined as:

$$\Phi(x) = \frac{1}{\sqrt{2\pi}} \int_{-\infty}^{x} e^{-\frac{u^2}{2}} du. \quad (9)$$

The joint probability between Alice and Bob is:

$$P_{A,B}(a,b) = P_A(a) P_{B|A}(b|a), \quad (10)$$

with $P_A(a)$denoting the prior probability of Alice's symbol ($P_A(a) = 1/4$ under equiprobable signaling).

### C. Decision Regions and Erasure Events

Due to the multi-threshold decision structure of the proposed receiver, two intermediate regions correspond to erasures ($D_1$ and$D_2$ in Fig. 2), for which no symbol is assigned. These events are discarded during the sifting stage and therefore do not contribute directly to key generation. The probability that a received sample falls within a valid decision region is defined as the acceptance probability:

$$P_{acc} = P(b \in \{0, 1, 2, 3\}), \quad (11)$$

which characterizes the fraction of channel uses that produce valid symbol decisions. In this work, the mutual information $I(A;B)$ used in the KGR calculation is evaluated over the valid decision outcomes only, i.e., conditioned on accepted events. Therefore, the KGR reported in the figures represents the number of secret bits generated per accepted channel use, while erasures reduce the net rate through the factor $P_{acc}$.

In addition to the acceptance probability, another important system-level metric is the secret throughput, which represents the number of secret bits generated per unit time, which is given by:

$$R_{\text{sec}} = P_{acc} \times KGR \times R_s, \quad (12)$$

where $R_s$ is the symbol rate and KGR is the secret key bits per channel use and it denotes the conditional key-generation rate evaluated for accepted events only.

## 4. BEAM-SPLITTING ATTACK (BSA) IN MT-IMDD

### A. Attack Model Description

In an IMDD–based SKD system such as the proposed MT-IMDD protocol, Eve attempts to infer the transmitted symbols by measuring the split signal without otherwise disturbing the optical field (see Fig. 3). Because the attack is passive, her presence does not introduce additional errors in Bob's measurements. However, Eve's information is fundamentally limited by the nonorthogonality of the underlying coherent states and by the quantum and classical noise inherent to direct-detection receivers.

In this attack model, Eve captures a fraction $(1 - T)$ of Alice's transmitted intensity $I_a$, corresponding to a symbol $a$ in the proposed MT-IMDD scheme. It is assumed that Eve is using an ideal quantum receiver ($\eta_E = 1$) and there is no thermal noise in her receiver. The current of Eve's receiver is:

$$i_{E,a} = \kappa(1 - T) I_a + n_E, \quad (13)$$

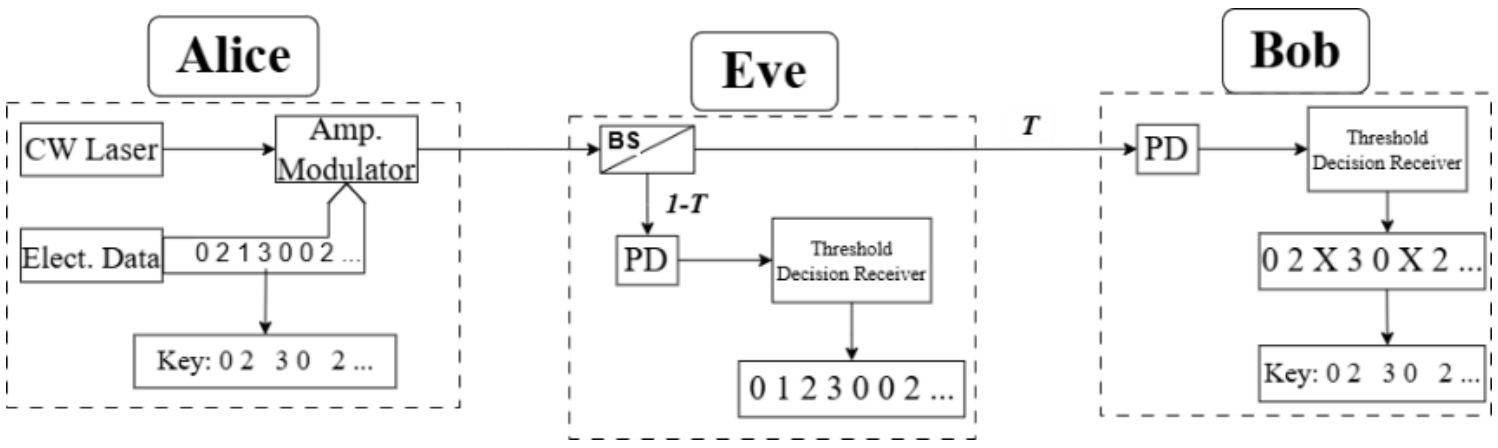

**Fig. 3.** Schematic of the proposed multi-threshold IMDD (MT-IMDD) protocol under a beam-splitting attack (BSA). Eve inserts a beam splitter (BS) to divert a fraction $1-T$ of the optical signal, detects it with her photodiode (PD) and threshold-based receiver, and attempts to infer Alice's symbols. Bob receives the remaining fraction $T$ and performs threshold detection to reconstruct his raw key.

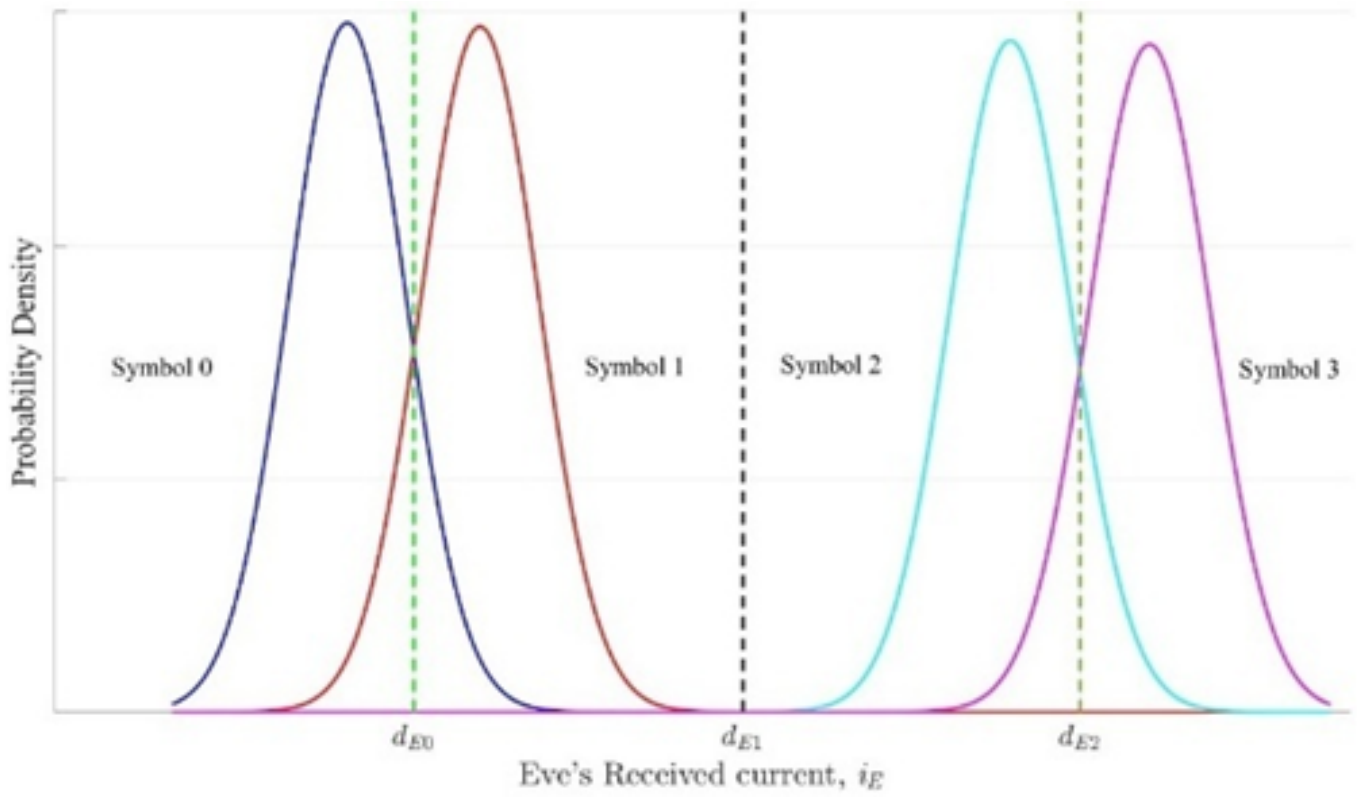

**Fig. 4.** Eve's received-current probability densities for the transmitted symbols in the proposed MT-IMDD protocol. Solid curves show the conditional Gaussian distributions corresponding to each symbol. The dashed vertical lines denote Eve's decision thresholds $d_{E0}, d_{E1}, d_{E2}$, which partition her measurement axis into symbol regions $\{0, 1, 2, 3\}$.

where $n_E$ represents the noise in Eve's measurement. The mean of $n_E$ is zero and its variance is:

$$\sigma_{E,a}^2 = \kappa^2(\sigma_{C,E,a}^2 + \sigma_{Q,E,a}^2),\ a = 0, 1, ..., 3 \tag{14}$$

where $\sigma_{C,E,a}^2$ is the variance of classical noise:

$$\sigma_{C,E,a}^2 = k_1 B(1-T)^2 I_a^2, \tag{15}$$

and $\sigma_{Q,E,a}^2$ is the variance of quantum noise:

$$\sigma_{Q,E,a}^2 = k_2 B(1-T) I_a. \tag{16}$$

Eve's ability to discriminate the transmitted symbols depends on the spacing between the intensity levels in the MT-IMDD alphabet and the magnitude of her detection noise. As shown in Fig. 4, she sets three thresholds $d_{E,e},\ e = 0, 1, 2$. When the received current is less than $d_{E,0}$, she creates symbol 0. If the current is in the range of $\left[d_{E,0}, d_{E,1}\right]$, she creates symbol 1, if the current is in the range of $\left[d_{E,1}, d_{E,2}\right]$, she creates symbol 2, and if the current exceeds $d_{E,2}$, she creates symbol 3.

In this study, Eve's receiver is modeled as using threshold-based detection consistent with the IMDD architecture assumed for all parties. Direct-detection photodiodes inherently produce noisy analog photocurrents that are ultimately converted into discrete symbol decisions through thresholding, and our analysis focuses on security within this discrete-modulation, multi-threshold framework. We note that prior works [25,26] have shown that an eavesdropper who retains and processes the full analog photocurrent, rather than applying a hard decision, can extract more information from the same IMDD signal. Incorporating such a soft-decision model for Eve would therefore increase her accessible information compared to a thresholded receiver. However, because the proposed MT-IMDD protocol is designed as a discrete-variable SKD scheme with symbol-level decision regions and erasures, our analysis evaluates security under the same class of measurements used to generate the raw key. A full soft-decision treatment for Eve, analogous to [25,26], constitutes a natural extension and would slightly decrease the key rate, but under realistic noise levels and nonorthogonal intensity states, the additional information gained is expected to be modest. Thus, the present model provides a representative and practically relevant estimate of Eve's information for discrete-level IMDD implementations.

Regarding the possibility of coherent detection with a local oscillator, the proposed MT-IMDD protocol encodes information solely in the optical intensity rather than in the optical phase. Consequently, coherent detection does not provide additional phase information beyond the intensity statistics already available through direct detection. As a result, the beam-splitting attack considered here captures the dominant leakage mechanism for intensity-modulated channels.

Eve's decision thresholds $d_{E,e}$ are so chosen that the mutual information $I(A;E)$ and $I(B;E)$ are maximized depending on forward reconciliation (FR) or reverse reconciliation (RR), respectively. The following sections evaluate Eve's mutual information and the corresponding key-generation rates based on this practically motivated model.

### B. Forward Reconciliation under Beam-Splitting Attack

In forward reconciliation (FR), Bob aligns his key with Alice's using classical error-correcting codes, with Alice's raw se-quence serving as the reference. The key-generation rate (KGR) under a BSA is:

$$KGR_{FR} = I(A;B) - I(A;E), \quad (17)$$

where $I(A;B)$ is evaluated using the joint probabilities of Eq. (10), restricted to the accepted decision outcomes($b \in \{0,1,2,3\}$), and $I(A;E)$ is obtained from Eve's Gaussian model of Section 4.A.

The expressions for the KGR used in this work correspond to the asymptotic information-theoretic secrecy bounds derived from the difference between the mutual information of the legitimate users and that of the eavesdropper. In practical secret-key distribution (SKD) systems, additional factors such as the reconciliation efficiency $\beta \le 1$, the error-correction leakage $Leak_{EC}$, and the cost of privacy amplification reduce the achievable key rate. Consequently, the rates reported in this work should be interpreted as upper bounds that characterize the fundamental performance of the proposed MT-IMDD protocol under ideal reconciliation conditions. In a practical implementation, the achievable secret rate can be approximated as $R = \beta I(A;B) - I(A;E) - Leak_{EC}$, for the forward-reconciliation case, with an analogous expression for reverse reconciliation. Typical reconciliation efficiencies in optical SKD systems are in the range $\beta \approx 0.9 - 0.95$. These practical factors scale the achievable rate but do not alter the relative performance trends demonstrated in the present study.

For each symbol $a$, Eve's received current follows a Gaussian pdf with mean $i_{E,a}$ and variance $\sigma^2_{E,a}$ given by Eqs. (13)–(16). The conditional probability that Eve assigns symbol $e$ is:

$$P_{E|A}(e|a) = \Phi\left(\frac{d_{E,e} - i_{E,a}}{\sigma_{E,a}}\right) - \Phi\left(\frac{d_{E,e-1} - i_{E,a}}{\sigma_{E,a}}\right), \; e = 0,1,2,3 \quad (18)$$

where $d_{E,-1} = -\infty$, $d_{E,3} = +\infty$.With equiprobable signaling $P_A(a) = 1/4$, the joint distribution is:

$$P_{A,E}(a,e) = P_A(a)P_{E|A}(e|a), \quad (19)$$

and $I(A;E)$ follows from Eq. (4). Finally, the KGR is calculated using Eq. (17).

### C. Reverse Reconciliation under Beam-Splitting Attack

In reverse reconciliation (RR), Alice aligns her key to match Bob's, treating Bob's raw key as the reference. The KGR is expressed as:

$$KGR_{RR} = I(A;B) - I(B;E). \quad (20)$$

Because Bob and Eve both observe a noisy version of Alice's transmitted data (but with correlated classical noise), computing $I(B;E)$ requires estimating their joint symbol probabilities:

$$P_{B,E}(b,e) = \sum_{a=0}^{3} P_{B,E|A}(b,e|a)P_A(a), \quad (21)$$

assuming uniform symbol probabilities $P_A(a) = 1/4$, this becomes:

$$P_{B,E}(b,e) = \frac{1}{4}\sum_{a=0}^{3} P_{B,E|A}(b,e|a). \quad (22)$$

When Alice sends a symbol of mean intensity $I_a$, the instantaneous emitted optical intensity fluctuates according to classical noise:

$$p_{A,C}(I_A|a) = \frac{1}{\sqrt{2\pi\sigma^2_{C,A,a}}}\exp\left[-\frac{(I_A - I_a)^2}{2\sigma^2_{C,A,a}}\right], \quad (23)$$

where $\sigma^2_{C,A,a} = k_1 B I_a^2$ represents the variance of classical intensity noise. When Alice transmits a symbol of mean intensity $I_a$, Bob and Eve each measure a noisy photocurrent following Gaussian distributions with respective means $i_{B,a} = \kappa\eta T I_a$ and $i_{E,a} = \kappa(1-T)I_a$, the corresponding noise variances are:

$$\begin{aligned} \sigma^2_{B,a} &= \kappa^2\sigma^2_{Q,B,a} + \sigma^2_T, \\ \sigma^2_{E,a} &= \kappa^2\sigma^2_{Q,E,a}. \end{aligned} \quad (24)$$

The conditional probability that Bob (or Eve) decides symbol $b$(or $e$) is obtained by integrating the corresponding pdf over the decision region bounded by their thresholds. Using the Gaussian cumulative distribution function $\Phi(x)$, these can be expressed as:

$$\begin{aligned} P_{B|A}(b_j|I_a) &= \Phi\left(\frac{d_{B,j} - i_{B,a}}{\sigma_{B,a}}\right) - \Phi\left(\frac{d_{B,j-1} - i_{B,a}}{\sigma_{B,a}}\right), \quad j = 0,1,...,5 \\ P_{E|A}(e|I_a) &= \Phi\left(\frac{d_{E,e} - i_{E,a}}{\sigma_{E,a}}\right) - \Phi\left(\frac{d_{E,e-1} - i_{E,a}}{\sigma_{E,a}}\right), \quad e = 0,1,2,3. \end{aligned} \quad (25)$$

The conditional joint probability that Bob and Eve obtain outcomes $b$ and $e$ when Alice transmits symbol $a$ is then:

$$P_{B,E|A}(b,e|a) = \int_0^{\infty} P_{B|A}(b|I_A)P_{E|A}(e|I_A)p_{A,C}(I_A|a)dI_A \;. \quad (26)$$

Finally, the marginal joint probabilities $P_{B,E}(b,e)$ required in Eq. (22) are obtained by summing over all possible symbols transmitted by Alice. Once $P_{B,E}(b,e)$ is computed, the mutual information $I(B;E)$ follows from Eq. (4), and the reverse reconciliation KGR is evaluated using Eq. (20).

More general eavesdropping strategies, such as collective attacks in which the eavesdropper performs joint processing over multiple signals, may provide higher information gain and require a quantum-mechanical analysis based on Holevo information. Such an analysis is beyond the scope of this work. It is worth noting, however, that in the high photon-number regime considered here, the detection statistics are well approximated by classical models, and the gap between Shannon's theory and Holevo-based bounds is expected to be reduced.

### D. Numerical Results

We analyze the joint optimization of modulation depth ($\delta$) and decision thresholds that lead to the maximum KGR. For this study, the symbol-0 intensity is 2 mW, and the modulation-depth spacing between symbols 1 and 2 is $\Delta_{12} = (I_2 - I_1)/I_1$. is held constant at 5% of their mean level. This parameter controls the separation between adjacent intensity levels (symbols 1 and 2) relative to the baseline intensity, and therefore directly determines the degree of overlap between the corresponding photocurrent distributions.

These parameters establish a representative operating point where both noise and attenuation are non-negligible. In the following results, the performance of the proposed multi-level encoding is compared with the binary IMDD scheme reported in [21], which serves as a baseline reference.

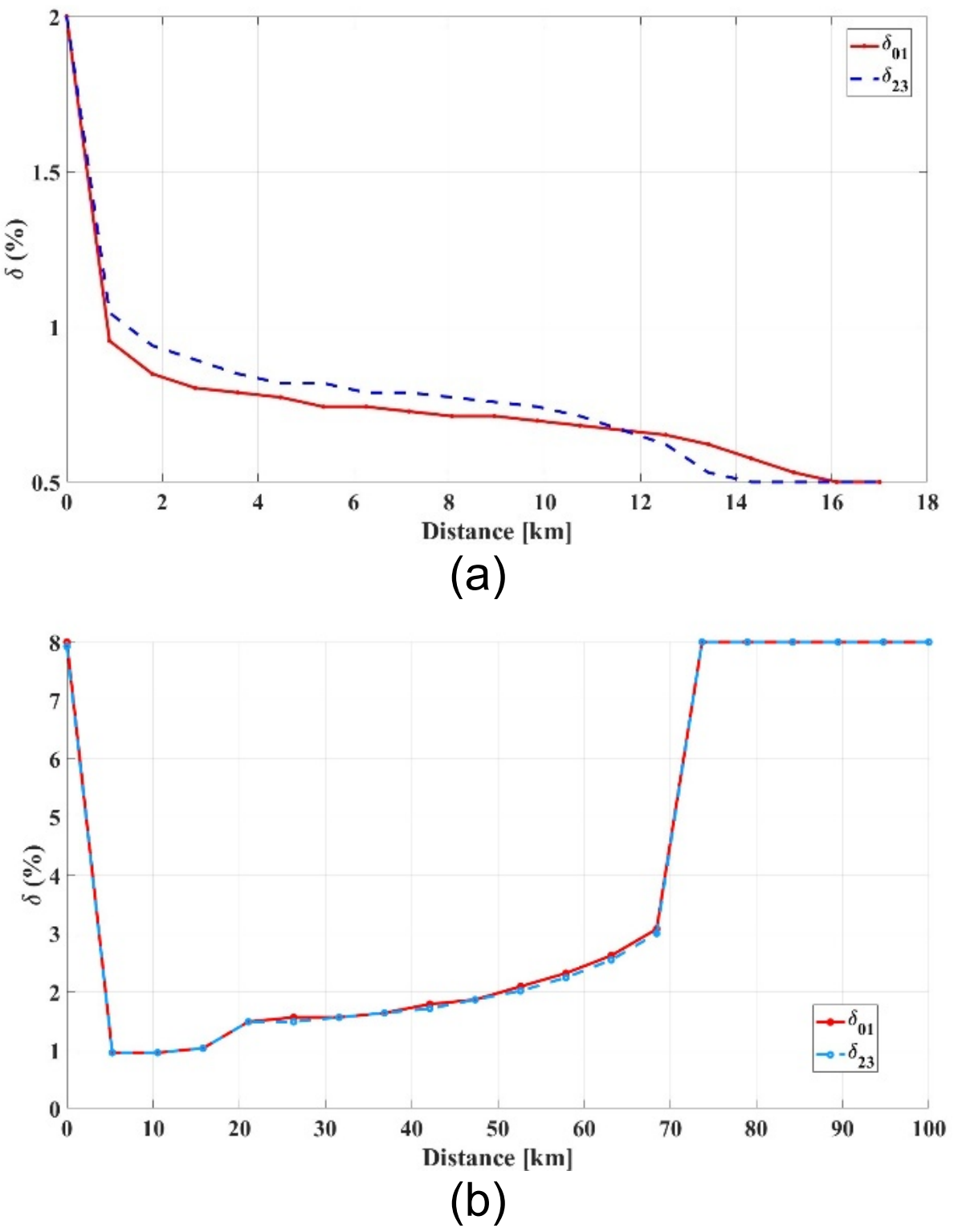

**Fig. 5.** Modulation depth that maximizes KGR versus distance. (a) Forward Reconciliation (FR). (b) Reverse Reconciliation (RR). The red (solid) and blue (dashed) curves correspond to $\delta_{01}$ and $\delta_{23}$, respectively.

The main simulation parameters used in the numerical analysis are summarized in Table 1. The optical power levels considered in the simulations remain in the milliwatt range (e.g., $I_0 = 2$ mW), which is well within the linear operating region of typical IM-DD transmitters and photodetectors; therefore, both the modulator and receiver are assumed to operate in their linear regimes over the full simulated distance range.

Figure 5 shows the modulation depth ($\delta$) that maximizes KGR as a function of distance for FR and RR. The two curves, $\delta_{01}$ ($= \delta_1$) and $\delta_{23}(= \delta_3 - \delta_2)$, correspond to the optimized intensity separations of the lower- and higher-power symbol pairs, respectively.

In FR (Fig. 5(a)), δ decreases rapidly with distance, large δ improves Bob's discrimination at short range but increases Eve's distinguishability at longer distances. The non-monotonic variation of the optimal modulation depth shown in Fig. 5(b) arises from the behavior of the $I_{BE}$ term in the $\mathrm{KGR}_{RR}$ expression, Eq. (20). Because Bob's and Eve's photocurrents scale with the transmittance $T$ and $1 - T$, respectively, their mutual information $I_{BE}$ exhibits a peak at intermediate distances. The rise of $I_{BE}$ at short to intermediate distances can be understood from the fact that their mutual information depends on the correlation between Bob's and Eve's received photocurrents. When both receivers obtain a strong optical signal, i.e., when the transmittance $T$and the tapped fraction $1 - T$are both appreciable, their detection outcomes become highly correlated, and $I_{BE}$ attains its maximum value. Conversely, when one receiver dominates (very high or very low $T$), the other observes a much weaker signal, reducing their statistical dependence and thus lowering $I_{BE}$. Around the distance where both Bob and Eve receive comparable optical powers (≈ 5 km in our simulation), $I_{BE}$ becomes the largest. At this point, although $I_{AB}$ remains high, our optimization algorithm selects a smaller modulation depth (δ) to suppress Eve's information gain, since the primary objective is minimizing $I_{BE}$. As distance increases further, $I_{BE}$ decreases because both receivers' signals weaken and are corrupted by stronger noise. Consequently, larger δ values become beneficial to recover Bob's distinguishability without substantially increasing Eve's knowledge, leading to the observed rise and eventual saturation of δ at longer distances.

In addition, heatmaps in Fig. 6 depict the KGR as a function of δ and Bob's decision thresholds ($d_{B0}$, $d_{B3}$). The white cross marks the optimal combination of δ and threshold that yields the maximum KGR. In FR (Fig. 6(a)), the optimum occurs near $\delta \approx 0.7\%$ and $d_{B0} \approx 9.72 \times 10^{-4}$ A, while in RR (Fig. 6(b)) it shifts to smaller δ and slightly higher threshold values. The RR contour exhibits a broader plateau, signifying greater robustness to threshold variations compared with FR.

The shapes of the KGR surfaces in Fig. 6 can be understood from the interplay between symbol separability and receiver noise. Increasing the modulation depth (δ) initially enhances Bob's ability to discriminate between neighboring symbols, but beyond the optimum, the expanded spacing also improves Eve's distinguishability and exposes Bob's outer thresholds to stronger noise tails, reducing secrecy. The optimal threshold values correspond closely to the intersection points of adjacent Gaussian distributions, where Bob's total decision error is minimized and $I_{AB}$ reaches its peak. The ridge-like high-KGR region observed around $\delta \approx 1\%$ for FR and slightly smaller δ for RR thus represents the balance between discrimination and secrecy. The broader high-KGR plateau seen in RR indicates that once Eve becomes noise-limited, the key rate becomes less sensitive to moderate variations in δ or threshold, highlighting RR's robustness. Furthermore, pair ($\delta_{23}$, $d_{B3}$) exhibits a wider and shallower optimum region than the pair ($\delta_{01}$, $d_{B0}$), reflecting its stronger photocurrents and reduced relative noise influence. These trends collectively confirm that the achievable KGR is governed by a delicate trade-off among optical power distribution, noise variance, and symbol-level separability.

Overall, these findings confirm that simultaneous optimization of the modulation depth (δ) and decision thresholds is essential for maximizing the secure key rate. In FR, the optimal δ decreases monotonically with distance, as smaller modulation depths help suppress Eve's information in high-loss regimes where Bob's signal-to-noise ratio remains relatively strong. In contrast, RR exhibits a non-monotonic δ behavior: δ first decreases as $I_{BE}$ rises to its mid-range maximum, then increases once $I_{BE}$ begins to decline at longer distances. This difference reflects the distinct optimization priorities of the two reconciliation strategies, FR prioritizing secrecy through reduced symbol separation, and RR maintaining efficiency by restoring Bob's distinguishability when Eve's correlation weakens.

Figure 7 shows how the achievable RR-KGR depends on the spacing between the two central intensity levels of the 4-level constellation, quantified by $\Delta_{12} = (I_2 - I_1)/I_1$. This parameter controls how far apart the lower and upper portions of the 4-symbol intensity structure are placed. When the separation is large (e.g., $\Delta_{12} \geq 5\%$), the four decision regions remain well defined, and the modulation behaves as a stable 4-level scheme, achieving a KGR close to its ideal performance. However, when $\Delta_{12}$ is reduced to values comparable to the modulation depth inside each cluster (≈1–3%), the intermediate levels $I_1$ and $I_2$ begin to overlap, which would increase the chance of detecting the transmitted symbols 2 and 3 as symbols 0 and 1, and vice versa. This overlap disrupts the reliability of symbol discrimination and simultaneously reduces Bob's information advantage over Eve, causing a steep decline in the secure key rate. In the limit of $\Delta_{12}$ approaching 0, Bob will not be able to distinguish between symbols 1 and 2, and the KGR drops to zero. The results highlight that maintaining sufficient separation between the central intensity levels is critical for preserving the performance of MT-IMDD secret-key distribution.

**Table 1. Simulation Parameters**

| Parameter | Value | Unit | Notes |
|---|---|---|---|
| Fiber attenuation | 0.2 | dB/km | Standard single-mode fiber loss |
| Receiver bandwidth | $9.5 \times 10^{9}$ | Hz | Photodiode electrical bandwidth |
| Detector quantum efficiency | 0.62 | — | Photodetector efficiency |
| Classical noise coefficient, $k_1$ | $5.66 \times 10^{-16}$ | — | Classical noise constant |
| Quantum noise coefficient, $k_2$ | $2.3 \times 10^{-19}$ | — | Shot-noise constant |
| Thermal noise coefficient, $k_3$ | $4.2 \times 10^{-23}$ | — | Thermal noise constant |
| Symbol rate | $10^{9}$ | Symbols/s | Used for throughput calculation |
| Base optical intensity | 2 | mW | Lowest symbol intensity level |

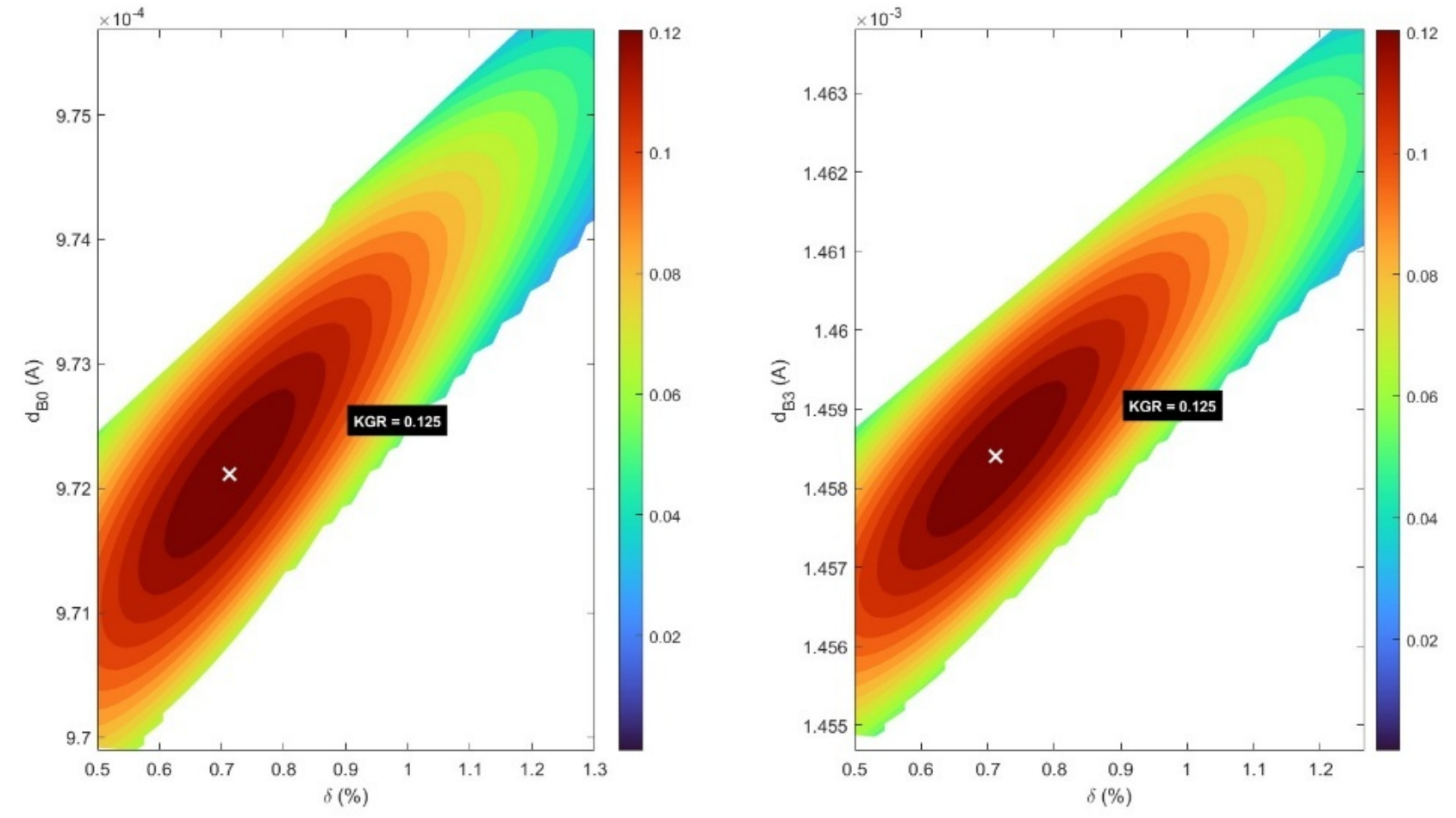

(a)

**Fig. 6.** Contour maps of the key-generation rate (KGR) as a function of modulation depth (δ) and Bob's decision thresholds (symbol-0 intensity = 2 mW, $\delta_{12}$ = 5%). (a) Forward Reconciliation (FR) at 10 km. (b) Reverse reconciliation (RR) at 80 km. The left panels correspond to the joint optimization of $\delta_{01}$ and $d_{B0}$ for the symbol pair (0 and 1), while the right panels correspond to $\delta_{23}$ and $d_{B3}$ for the pair (2 and 3). The white marker denotes the optimum that maximizes the KGR.

Numerical simulation results of KGR for the two different reconciliation strategies are shown in Fig. 8. For each distance, we used the optimized modulation depth and decision thresholds. In the case of FR (Fig. 8(a)), when $T \cong 1$ (i.e. distance ≅ 0 km), we see that the KGR for our new scheme (MT-IMDD) is enhanced by a factor of ~2, consistent with the fact that in the absence of noise, the Shannon capacity of the proposed protocol (MT-IMDD) is 2 bits/symbol, whereas that of Ikuta-Inoue's scheme is only 1 bit/symbol. In the case of FR, when the transmission distance is 14 km ($T \cong 0.52$), the KGR is enhanced by a factor of 10.86 using MT-IMDD compared to Ikuta–Inoue's protocol (see Fig. 8(a)). In the case of RR (Fig. 8(b)), transmission distance is significantly enhanced compared to the case of FR for both schemes. As can be seen from Fig. 8(b), KGR for MT-IMDD is higher than Ikuta–Inoue's protocol over the range of transmission distances studied. When the distance = 100 km, the KGR of MT-IMDD is enhanced by a factor of 13.5, compared to Ikuta–Inoue's scheme.

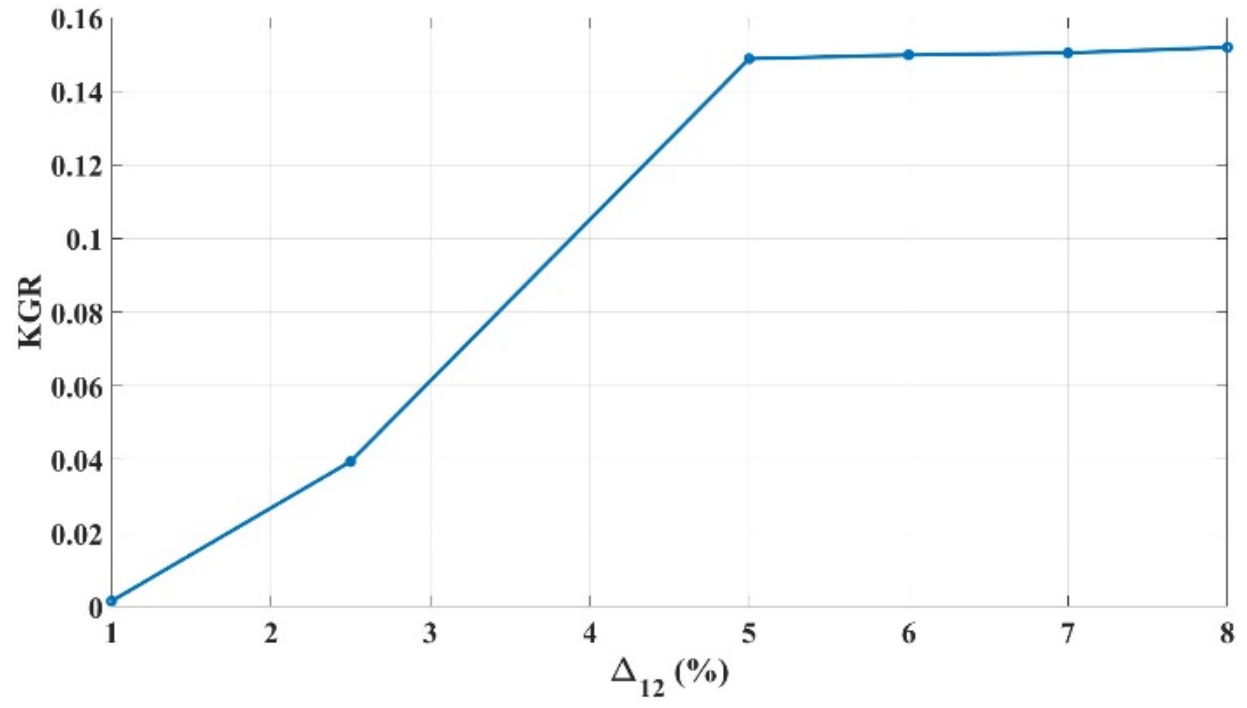

**Fig. 7.** Peak total RR key-generation rate (KGR) as a function of the inter-subset intensity spacing $\Delta_{12} = (I_2 - I_1)/I_1$ at 80 km. For each value of $\Delta_{12}$, the KGR is maximized over the intra-pair modulation depth $\delta$ and decision threshold $d_{B0}$ ($d_{B3}$). When the symbol subsets are well separated ($\Delta_{12} \geq 5\%$), the two binary sub-channels are effectively independent, and both achieve a KGR near their ideal limit. As $\Delta_{12}$ is reduced to values comparable to the intra-pair modulation ($\Delta_{12} = 1\% - 3\%$), the overlap between the intermediate levels $I_1$ and $I_2$ causes strong cross-talk between the two neighbor symbols, resulting in a dramatic reduction of the total achievable KGR.

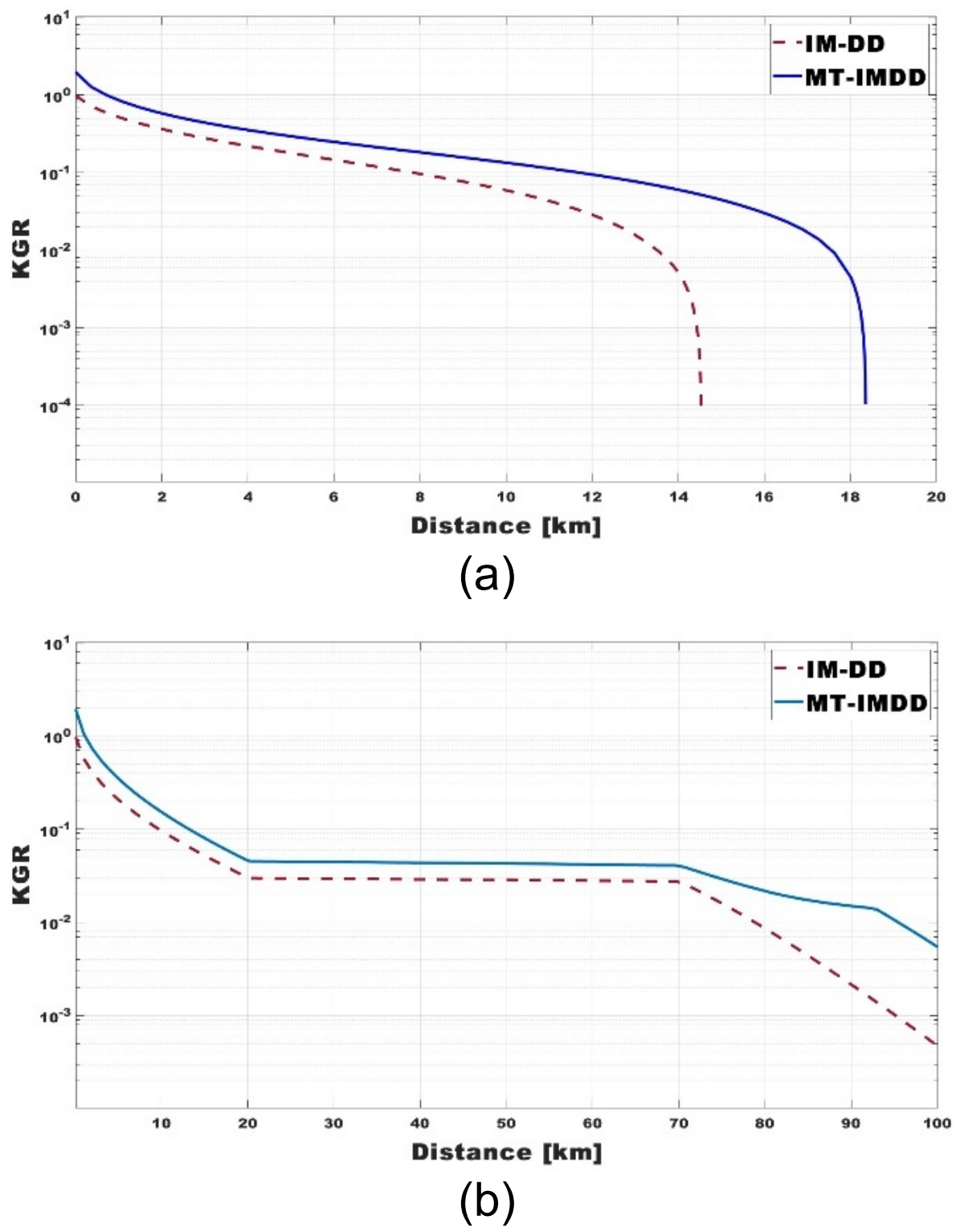

**Fig. 8.** Key-generation rate (KGR) versus transmission distance under a beam-splitting attack for forward reconciliation (FR) and reverse reconciliation (RR) with optimized modulation depths. Results are shown for binary IMDD (red dashed curves) and the proposed multi-threshold IMDD (MT-IMDD) scheme (solid curves). Subfigure (a) corresponds to FR, where MT-IMDD provides a substantial KGR enhancement at short and moderate distances and extends the maximum secure distance from approximately 14 km (binary) to about 18 km. Subfigure (b) shows RR, which significantly increases the secure transmission range for both schemes. Across the entire RR distance range, MT-IMDD consistently achieves higher KGR than Ikuta-Inoue's scheme, with especially pronounced gains in the first 20–40 km region.

In addition to the KGR, it is also instructive to examine the acceptance probability and the resulting secret throughput, which together characterize the practical operating trade-off of the receiver. As discussed in Section 3.B, erasure events occur when the received photocurrent falls within the intermediate decision regions and no symbol decision is assigned. The acceptance probability therefore represents the fraction of channel uses that lead to valid symbol decisions, while the conditional KGR represents the number of secret bits extracted per accepted event.

Figure 9 illustrates the relationship between acceptance probability, KGR, and secret throughput at a transmission distance of approximately 10 km for the BSA receiver under FR. The curves are obtained by sweeping the modulation depth and decision thresholds, which changes the size of the erasure regions and the overlap between the conditional photocurrent distributions. Increasing the separation between decision thresholds reduces the probability of symbol error among accepted events, thereby improving the conditional KGR. However, this also enlarges the erasure regions $D_1$ and $D_2$, which reduces the acceptance probability because a larger fraction of received samples fall into the intermediate decision intervals where no symbol is assigned.

As a result, the KGR increases as the acceptance probability decreases, reflecting the trade-off between reliability and the fraction of retained symbols. The secret throughput exhibits a non-monotonic dependence on the acceptance probability: when the acceptance probability is close to unity, the symbol decisions are more frequent but less reliable, leading to lower mutual information. Conversely, when the acceptance probability becomes very small, only highly reliable symbols are retained, but the number of accepted events is insufficient to sustain a high key rate. The maximum secret throughput therefore occurs at an intermediate acceptance probability that balances decision reliability and symbol utilization.

These results highlight that although the KGR quantifies the amount of secret information per accepted symbol, the effective system performance must also account for the probability of acceptance. The secret throughput therefore reflects the combined impact of mutual information and erasure filtering, providing a more complete measure of practical key generation performance under the BSA with FR.

To provide further physical insight into the detection process, we evaluate the SER and the acceptance probabilities associated with the decision regions outside the erasure intervals $D_1$ and $D_2$ as functions of transmission distance, as shown in Fig. 10. These results correspond to the optimized BSA under FR, where the decision thresholds are selected to maximize the KGR at each distance. The SER is computed only over accepted events, while the acceptance probability $P_{acc}$ represents the fraction of channel uses for which the received photocurrent falls outside the intermediate erasure regions and therefore produces a valid symbol decision.

As the transmission distance increases, fiber attenuation reduces the separation between the mean photocurrents corresponding to different intensity levels. The resulting increase in distribution overlap increases the probability of incorrect decisions among accepted symbols, leading to a gradual increase in the SER. At the same time, the reduced separation between adjacent photocurrent distributions enlarges the effective erasure regions $D_1$ and $D_2$, decreasing the probability that the received signal falls within the acceptance regions. Consequently, the overall acceptance probability decreases with distance.

First, let us focus on the erasure region $D_1$. When the transmission distance is 0 km, the decision thresholds $d_{B,0}$ and $d_{B,1}$ coincide and they will be at the middle of the first two peaks (see Fig. 2), and in this case, the erasure region $D_1$ is non-existent. As the distance increases, the decision thresholds move in the opposite directions and the erasure region is widened leading to lower acceptance probability. The acceptance probability associated with the region $D_2$ can be similarly explained.

These trends directly explain the behavior observed in Fig. 8(a), where the KGR decreases with distance due to the combined effect of increased decision errors and reduced distinguishability between symbols.

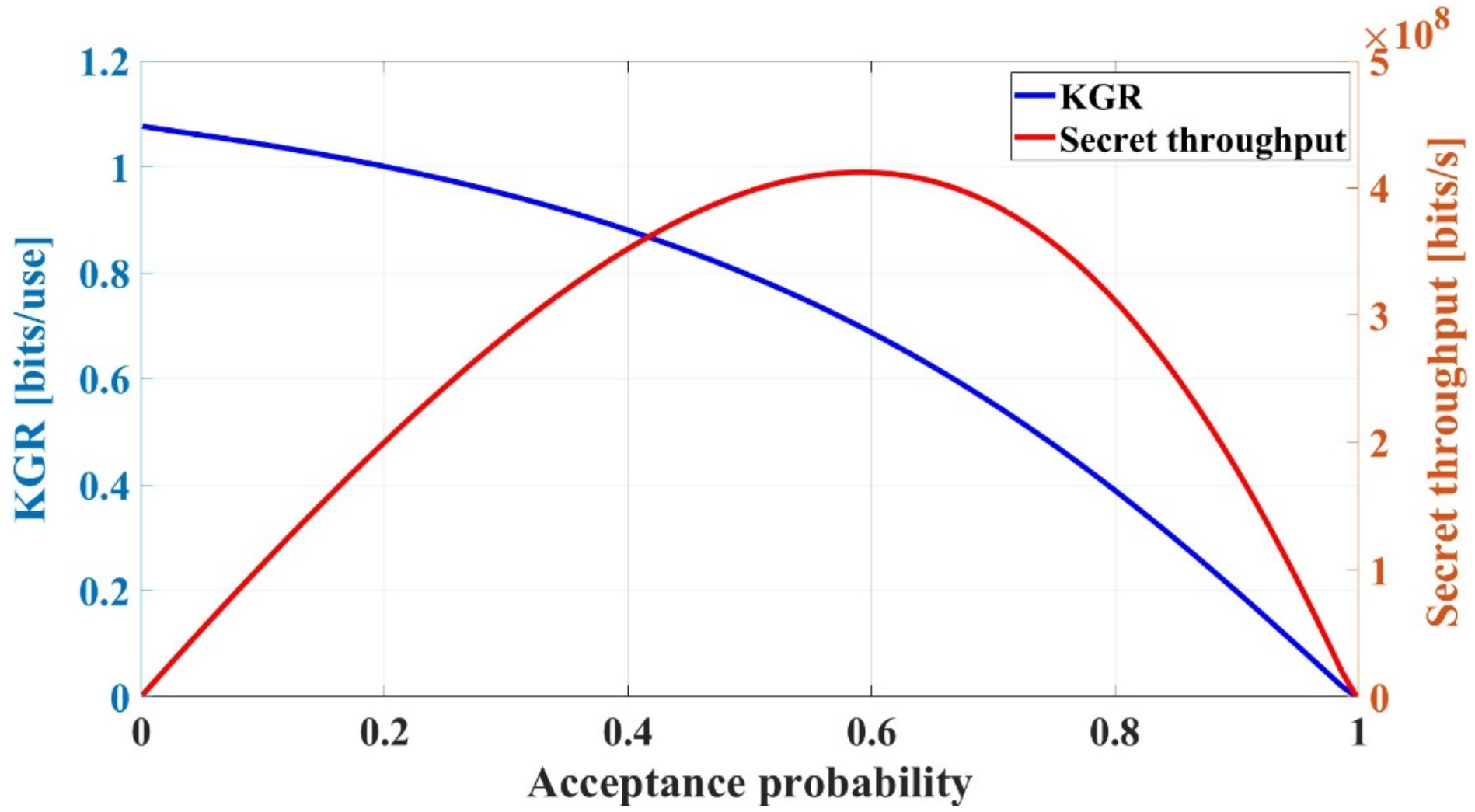

**Fig. 9.** KGR and secret throughput versus acceptance probability for the BSA receiver under FR at a transmission distance of approximately 10 km. Increasing the separation between decision thresholds reduces symbol errors among accepted events, improving the conditional KGR, but also enlarges the erasure regions $D_1$ and $D_2$, which decreases the acceptance probability. The secret throughput therefore exhibits an optimal operating point that balances reliability and the fraction of accepted symbols.

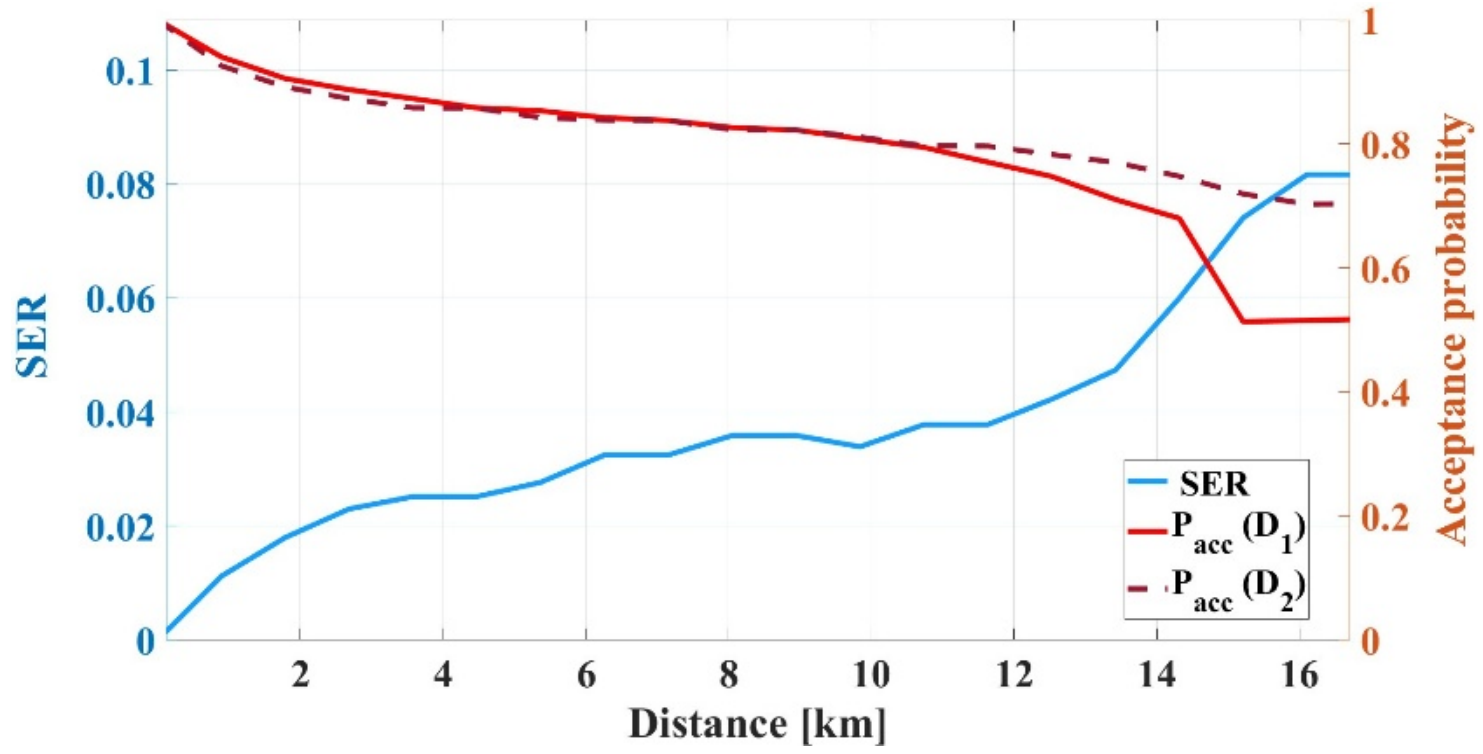

**Fig. 10.** SER and acceptance probabilities associated with the erasure regions $D_1$ and $D_2$ versus transmission distance for the BSA under FR. The blue solid curve shows the SER (left axis), while the red solid and red dashed curves represent the acceptance probabilities corresponding to decision regions outside $D_1$ and $D_2$, respectively (right axis). Increasing distance increases the overlap between symbol photocurrent distributions, resulting in higher SER and reduced acceptance probability due to enlarged erasure regions.

## 5. INTERCEPT–RESEND ATTACK (IRA) IN MT-IMDD

In this section, we analyze the impact of intercept–resend attacks (IRA) on the proposed multi-threshold IMDD (MT-IMDD) secret-key distribution scheme (Fig. 11). Rather than following the modeling approach of [19], we adopt a framework analogous to the standard IRA treatment used in discrete-variable protocols such as BB84 [25]. Under this model, whenever Eve intercepts and measures a transmitted symbol, the corresponding noise variance in her measurement is characterized as follows:

$$\begin{aligned} \sigma'^2_{E,a} &= B(k_1 i'^2_{E,a} + k_2 R i'_{E,a}),\ a = 0, 1, ..., 3, \\ i'_{E,a} &= \kappa I_{E,a}, a = 0, 1, ..., 3, \end{aligned} \quad (27)$$

where $I_{E,a}$ is the mean optical intensity at Eve's receiver when Alice sends symbol $a$. Eve resends the symbols to Bob based on her own measurements. For the part of the signal that Bob receives from Eve, the noise level increases because Eve's measurement introduces additional noise. However, for the part of the signal that Bob receives directly from Alice, the noise variance remains the same as before (Eq. (7)). The joint probability of Alice sending a symbol $a$ and Bob detecting a symbol $b$ is:

$$P_{A,B}(a,b) = xP_{A,B|att}(a,b|1) + (1-x)P_{A,B|att}(a,b|0), \quad (28)$$

where $x \equiv P_{att}(1)$ is the probability that Eve attacks. $P_{A,B|att}(a,b|0)$ represents the joint probability between Alice and Bob when Eve does not attack and it is calculated in Section 3.B. To calculate the joint probability between Alice and Bob when Eve attacks, we proceed as follows:

$$P_{A,B|att}(a,b|1) = \sum_{e=0}^{3} P_{E|A}(e|a)P_{B|E}(b|e)P_A(a). \quad (29)$$

Based on Eve's thresholds, conditional probability that Eve creates symbol $e$, $e \in \{0,1,2,3\}$, when Alice sends symbol $a$ is:

$$P_{E|A}(e|a) = \Phi\left(\frac{d_{E,e} - i'_{E,a}}{\sigma'_{E,a}}\right) - \Phi\left(\frac{d_{E,e-1} - i'_{E,a}}{\sigma'_{E,a}}\right). \quad (30)$$

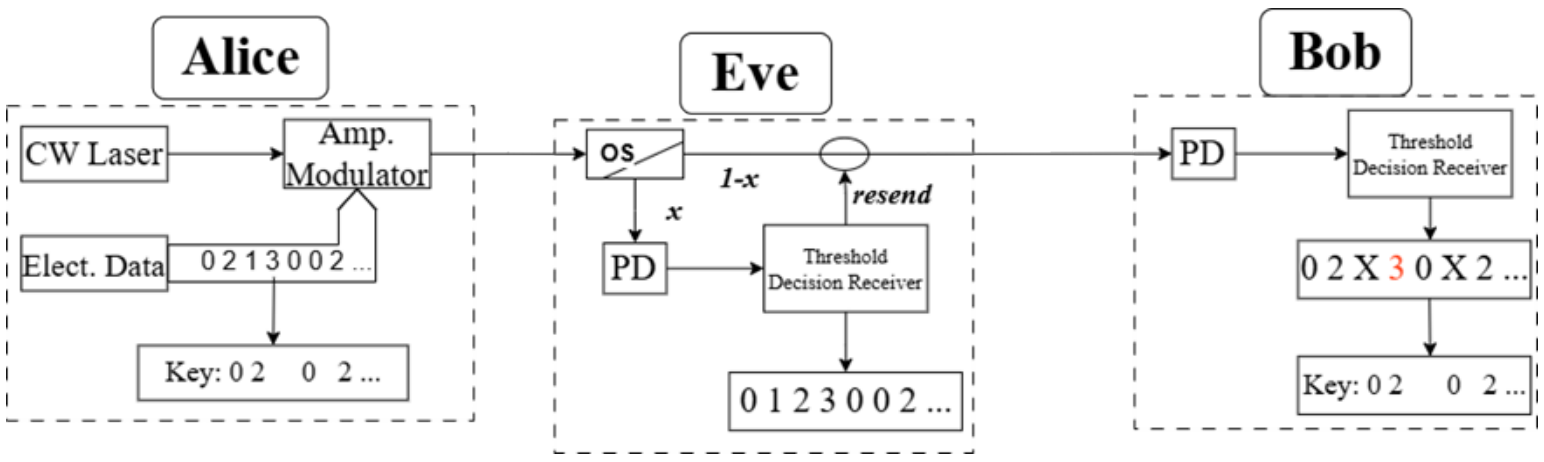

**Fig. 11.** Schematic of the proposed multi-threshold IMDD (MT-IMDD) protocol under an intercept–resend attack (IRA). Eve uses an optical switch (OS) to intercept a fraction $x$ of the transmitted signals, measures them with a photodiode (PD) and a threshold-decision receiver, and resends newly prepared symbols to Bob. Bob applies the same threshold-based detection to obtain his raw key.

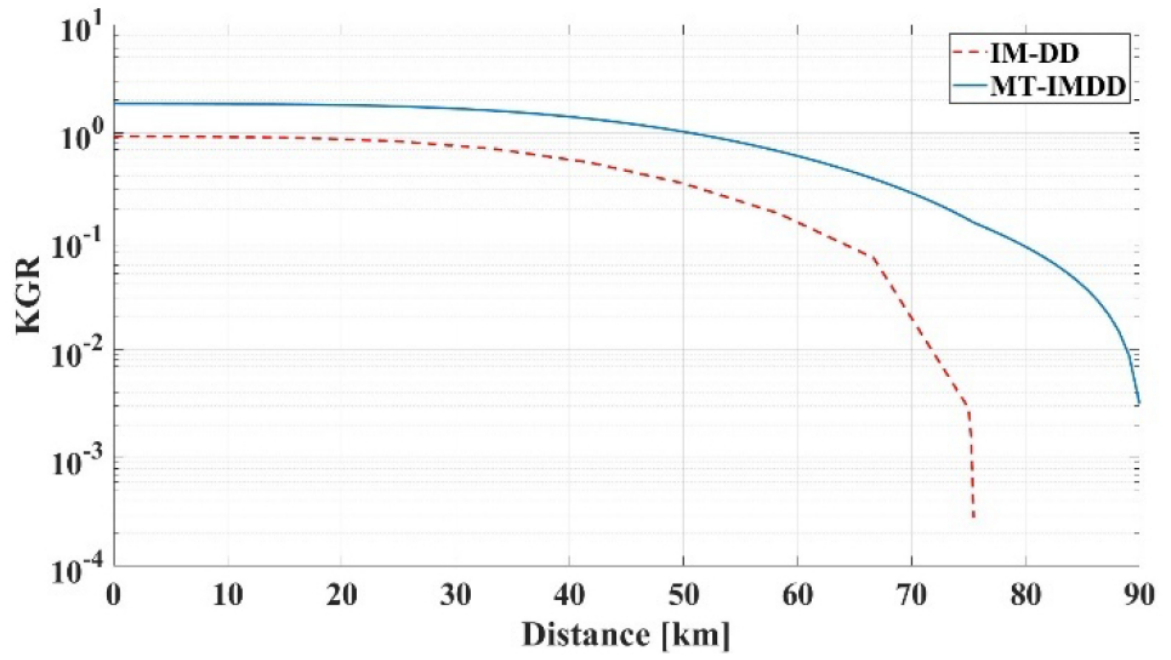

(a)

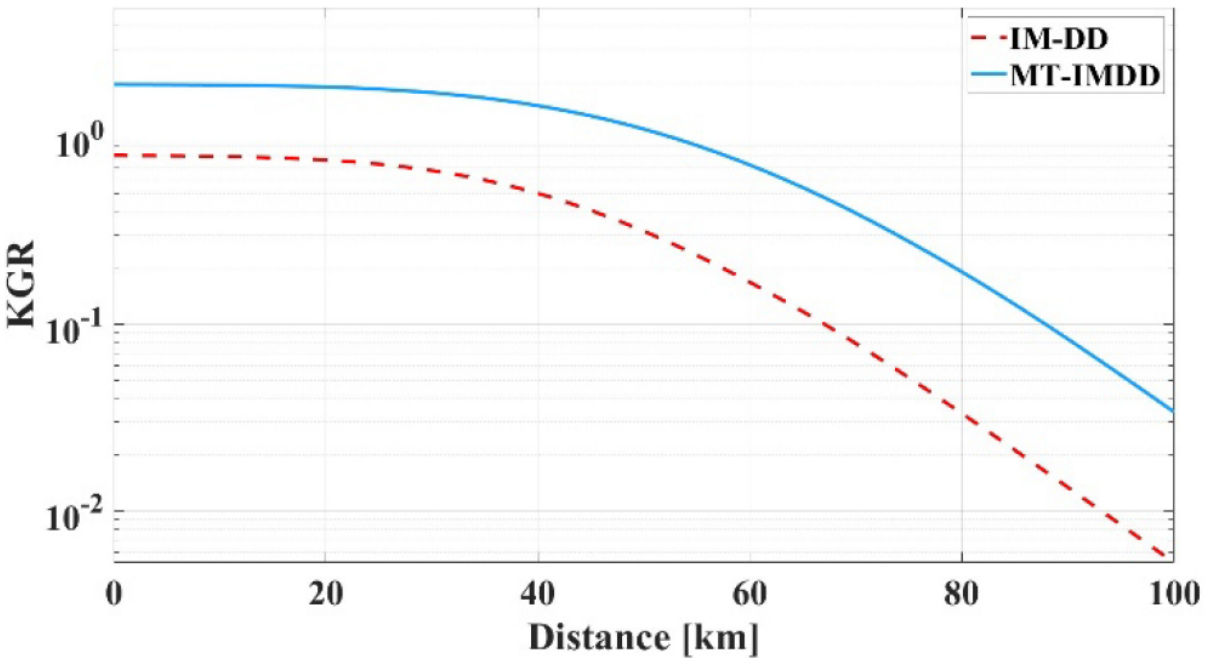

(b)

**Fig. 12.** Key Generation Rate (KGR) as a function of distance under the intercept–resend attack, comparing forward and reverse reconciliation schemes when $x = 0.2$. Ikuta–Inoue's protocol (red dashed lines) and the proposed protocol (blue solid lines) are shown for both cases. Subfigure (a) corresponds to forward reconciliation (FR), while subfigure (b) corresponds to reverse reconciliation (RR). The results demonstrate that RR significantly outperforms FR in high-loss (long-distance) regimes. Additionally, MT-IMDD shows a clear advantage over IMDD at short to moderate distances in both reconciliation approaches.

Similarly, the conditional probability that Bob creates symbol $b$, $b \in \{0, D_1, 1, 2, D_2, 3\}$, when Eve sends symbol $e$ is:

$$P_{B|E}(b_j|e) = \Phi\left(\frac{d_{B,j} - i_{B,e}}{\sigma_{B,e}}\right) - \Phi\left(\frac{d_{B,j-1} - i_{B,e}}{\sigma_{B,e}}\right), \quad j = 0, 1, ..., 5 \tag{31}$$

where $i_{B,e} = \kappa I_{B,e}$ and $\sigma_{B,e}$ denote the mean photocurrent and standard deviation at Bob's receiver for symbol $e$.

Using Eqs. (30) and (31) in (29), we can compute $P_{A,B|att}(a, b|1)$. For the non-attacked portion of the transmission $(1 - x \equiv P_{att}(0))$, Eve does not measure the signal but can still assign a random symbol. Therefore, the conditional joint probability $P_{A,E|att}(a, e|0)$ is $1/16$ for each $e$, since $P_A(a) = 1/4$ and Eve's choice is independent of $a$. Consequently, the overall joint probability between Alice and Eve becomes:

$$\begin{aligned} P_{A,E}(a, e) &= xP_{A,E|att}(a, e|1) + (1 - x)P_{A,E|att}(a, e|0) \\ &= xP_{E|A}(e|a)P_A(a) + \frac{1}{16}(1 - x). \end{aligned} \tag{32}$$

The mutual information $I(A; E)$ is calculated using Eqs. (4), (30), and (32). Similarly, the joint probability between Bob and Eve is:

$$\begin{aligned} P_{B,E}(b, e) &= xP_{B,E|att}(b, e|1) + (1 - x)P_{B,E|att}(b, e|0) \\ &= xP_{B,E|att}(b, e|1) + \frac{1}{16}(1 - x). \end{aligned} \tag{33}$$

The joint probability between Bob and Eve when Eve attacks is:

$$\begin{aligned} P_{B,E|att}(b, e|1) &= P_{B|E}(b|e)P_E(e), \\ P_E(e) &= \sum_{a=0}^{3} P_{E|A}(e|a)P_A(a). \end{aligned} \tag{34}$$

$P_{E|A}(e|a)$and $P_{B|E}(b|e)$ can be calculated using Eqs. (30) and (31), respectively. Using Eqs. (33) and (34) in Eq. (4), $I(B; E)$ is calculated. Finally, the KGR for FR and RR are Eqs. (17) and (20), respectively.

The simulation results of KGR for these two cases are plotted in Fig. 12. When the transmission distance is 50 km, the KGR is enhanced by a factor of 2.995 using the proposed scheme compared to Ikuta–Inoue's scheme for FR (Fig. 12(a)). In the case of RR (Fig. 12(b)), transmission distance is significantly enhanced compared to the case of FR for both schemes. As can be seen from Fig. 12(b), KGR for our protocol (MT-IMDD) is higher than Ikuta–Inoue's protocol over the range of transmission distances studied. When the distance = 100 km, the KGR of MT-IMDD is enhanced by a factor of 2.25, compared to IMDD.

To illustrate how Eve's attack intensity affects the achievable secrecy, Fig. 13 shows the KGR as a function of the attack probability $x$ at a fixed distance of 70 km. The blue and orange solid curves represent the forward- and reverse-reconciliation key rates ($K_{FR}$ and $K_{RR}$), respectively, while the dashed lines show the corresponding mutual-information terms $I(A; E)$, $I(B; E)$, and $I(A; B)$. As $x$ increases, both key rates drop nearly linearly because Eve's accessible information rises proportionally to her attack fraction, whereas the mutual information between Alice and Bob, $I_{AB}$, varies only slightly under the modest resend noise assumed. Reverse reconciliation maintains a positive KGR over a wider attack range, demonstrating its superior resilience against intercept–resend attacks (Fig. 13).

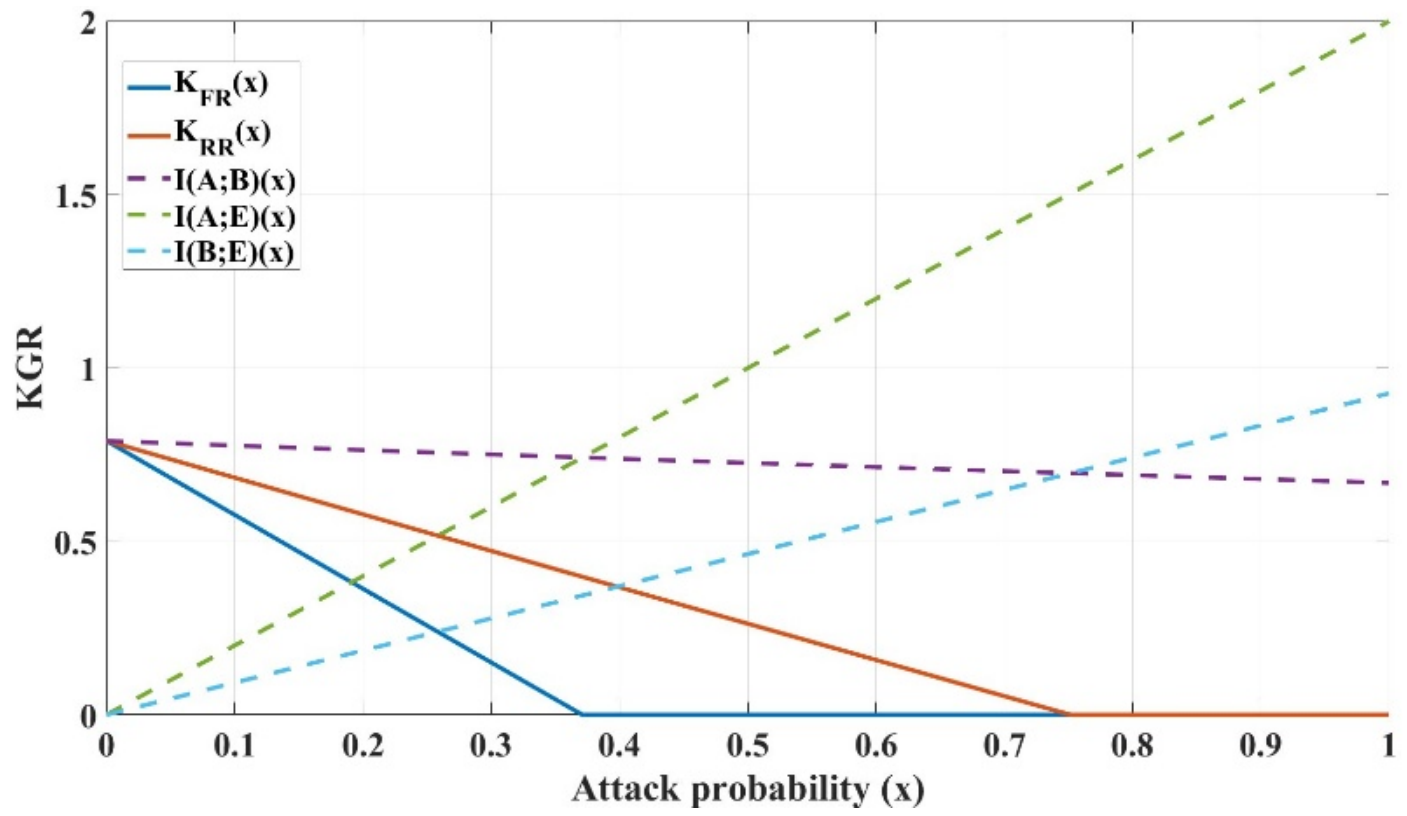

**Fig. 13.** KGR versus attack probability under intercept–resend attack at a distance of 70 km. The blue and orange solid lines show the $K_{FR}(x)$ and $K_{RR}(x)$, respectively. The purple, green, and cyan dashed lines correspond to $I(A;B)$, $I(A;E)$, and $I(B;E)$, respectively. Reverse reconciliation maintains a higher key rate, indicating greater resilience to active eavesdropping.

## 6. CONCLUSION

We have presented a new discrete-variable secret-key distribution (DV-SKD) protocol based on a multi-threshold intensity-modulated direct-detection (MT-IMDD) architecture. By employing a multi-symbol intensity alphabet together with optimized decision thresholds and erasure regions, the proposed scheme achieves higher key-generation rates (KGR) than binary IMDD across short to moderate transmission distances while maintaining full compatibility with existing optical communication hardware.

A central outcome of our analysis is the joint optimization of modulation depths and threshold locations, which balances symbol distinguishability for the legitimate users against information leakage to an eavesdropper. Numerical evaluations show that, even under substantial channel loss, reverse reconciliation enables the MT-IMDD scheme to maintain a positive secure-key rate and consistently outperform IMDD in both throughput and robustness. At a transmission distance of 100 km, our MT-IMDD protocol demonstrates a 13.5× and 2.25× improvement in the key-generation rate (KGR) compared to the Ikuta–Inoue scheme under reverse-reconciliation (RR) conditions, for the beam-splitting attack (BSA) and intercept-resend attack (IRA), respectively. Since the KGR is the primary performance metric in secure-key distribution (SKD), these enhancements clearly highlight the substantial advantage and practical impact of our proposed protocol.

Future work includes experimental validation of the MT-IMDD architecture under practical impairments such as finite extinction ratios, detector noise characteristics, and hardware nonlinearities. Extending the multi-threshold framework to higher-order intensity alphabets (e.g., 6- or 8-level constellations) may further improve spectral efficiency, though such extensions must be carefully assessed with respect to hardware complexity and security margins. These directions will inform the development of next-generation IMDD-based SKD systems that offer a compelling balance of practicality, scalability, and high secret-key throughput.

**Disclosures.** The authors declare that there are no conflicts of interest related to this article.

**Data availability.** Data underlying the results presented in this paper are not publicly available but may be obtained from corresponding author.

## REFERENCES

1. C. H. Bennett and G. Brassard, "An update on quantum cryptography," in Workshop on the Theory and Application of Cryptographic Techniques (Springer, 1984).
2. V. Scarani, H. Bechmann-Pasquinucci, N. J. Cerf, *et al.*, "The security of practical quantum key distribution," Rev. Modern Phys. **81**, 1301–1350 (2009).
3. C. H. Bennett, "Quantum cryptography using any two nonorthogonal states," Phys. Rev. Lett. **68**, 3121–3124 (1992).
4. A. K. Ekert, "Quantum cryptography based on Bell's theorem," Phys. Rev. Lett. **67**, 661–663 (1991).
5. D. Bruß, "Optimal eavesdropping in quantum cryptography with six states," Phys. Rev. Lett. **81**, 3018–3021 (1998).
6. M. Esmann, S. C. Wein, and C. Antón-Solanas, "Solid-state single-photon sources: recent advances for novel quantum materials," Adv. Funct. Mater. **34**, 2315936 (2024).
7. T. Attema, J. W. Bosman, and N. M. Neumann, "Optimizing the decoy-state BB84 QKD protocol parameters," Quantum Inf. Process. **20**, 154 (2021).
8. F. Grosshans and P. Grangier, "Continuous variable quantum cryptography using coherent states," Phys. Rev. Lett. **88**, 057902 (2002).
9. M. N. Notarnicola, M. Jarzyna, S. Olivares, *et al.*, "Optimizing state-discrimination receivers for continuous-variable quantum key distribution over a wiretap channel," New J. Phys. **25**, 103014 (2023).
10. F. Grosshans, G. Van Assche, J. Wenger, *et al.*, "Quantum key distribution using Gaussian-modulated coherent states," Nature **421**, 238–241 (2003).
11. N. J. Cerf, M. Levy, and G. Van Assche, "Quantum distribution of Gaussian keys using squeezed states," Phys. Rev. A **63**, 052311 (2001).
12. S. Pirandola, U. L. Andersen, L. Banchi, *et al.*, "Advances in quantum cryptography," Adv. Opt. Photon. **12**, 1012–1236 (2020).
13. C. Weedbrook, S. Pirandola, R. García-Patrón, *et al.*, "Gaussian quantum information," Rev. Modern Phys. **84**, 621–669 (2012).
14. Y. Zhang, Y. Bian, Z. Li, *et al.*, "Continuous-variable quantum key distribution system: past, present, and future," Appl. Phys. Rev. **11**, 011318 (2024).
15. T. C. Ralph, A. Gilchrist, G. J. Milburn, *et al.*, "Quantum computation with optical coherent states," Phys. Rev. A **68**, 042319 (2003).
16. M. Cattaneo, M. G. Paris, and S. Olivares, "Hybrid quantum key distribution using coherent states and photon-number-resolving detectors," Phys. Rev. A **98**, 012333 (2018).
17. N. Alshaer, M. E. Nasr, and T. Ismail, "Hybrid MPPM-BB84 quantum key distribution over FSO channel considering atmospheric turbulence and pointing errors," IEEE Photon. J. **13**, 1–9 (2021).
18. A. Yazdanpour, S. Kumar, M. Khajezadeh, *et al.*, "Enhancement of key generation rate in QKD systems using dual branch PNR detectors," Opt. Express **32**, 13583–13596 (2024).

19. A. Sanvito, S. Cassina, M. Lamperti, *et al.*, “Assessing a binary quantum channel exploiting a silicon photomultiplier based hybrid receiver,” Opt. Express **32**, 39846–39859 (2024).
20. M. T. DiMario and F. E. Becerra, “Robust measurement for the discrimination of binary coherent states,” Phys. Rev. Lett. **121**, 023603 (2018).
21. T. Ikuta and K. Inoue, “Intensity modulation and direct detection quantum key distribution based on quantum noise,” New J. Phys. **18**, 013018 (2016).
22. N. Yamamori and K. Inoue, “Experimental demonstration of intensity-modulation/direct-detection secret key distribution,” Jpn. J. Appl. Phys. **59**, 022003 (2020).
23. L. Mandel and E. Wolf, “Coherent states of the electromagnetic field,” in *Optical Coherence and Quantum Optics* (Cambridge University Press, 1995), pp. 522–572.
24. C. E. Shannon, “A mathematical theory of communication,” Bell Syst. Techn. J. **27**, 379–423 (1948).
25. T. A. Eriksson, P. V. Trinh, H. Endo, *et al.*, “Secret key rates for intensity-modulated dual-threshold detection key distribution under individual beam splitting attacks,” Opt. Express **26**, 20409–20419 (2018).
26. K. Banaszek, M. Jachura, P. Kolenderski, *et al.*, “Optimization of intensity-modulation/direct-detection optical key distribution under passive eavesdropping,” Opt. Express **29**, 43091–43103 (2021).